\documentclass[journal=ancac3,manuscript=article]{achemso}

\usepackage{amsmath}
\usepackage{amssymb}
\usepackage{bm}

\usepackage{graphicx}

\title{An All-Electric Single-Molecule Motor}

\author{Johannes~S.~Seldenthuis}
\email{j.s.seldenthuis@tudelft.nl}
\author{Ferry~Prins}
\author{Joseph~M.~Thijssen}
\author{Herre~S.~J.~van~der~Zant}
\affiliation{Kavli Institute of Nanoscience, Delft University of Technology, Lorentzweg 1, 2628 CJ Delft, The Netherlands}

\begin{document}

\begin{abstract}
Many types of molecular motors have been proposed and synthesized in recent years, displaying different kinds of motion, and fueled by different driving forces such as light, heat, or chemical reactions. We propose a new type of molecular motor based on electric field actuation and electric current detection of the rotational motion of a molecular dipole embedded in a three-terminal single-molecule device. The key aspect of this all-electronic design is the conjugated backbone of the molecule, which simultaneously provides the potential landscape of the rotor orientation and a real-time measure of that orientation through the modulation of the conductivity. Using quantum chemistry calculations, we show that this approach provides full control over the speed and continuity of motion, thereby combining electrical and mechanical control at the molecular level over a wide range of temperatures. Moreover, chemistry can be used to change all key parameters of the device, enabling a variety of new experiments on molecular motors.
\end{abstract}


Motivated by examples found in nature, and propelled by recent advances in synthetic chemistry, the field of molecular motors is rapidly developing into a major area of research \cite{Browne2006,Kay2007,Koumura1999,Kelly1999,Leigh2003,Zheng2004,Fletcher2005,Baber2008,Wang2008,Gao2008,Dundas2009,Ye2010}. Molecular motors are (supra-)molecules that are able to convert energy into continuous directional motion of one molecular component relative to another. For a molecule to be able to function as a motor, at least two (meta)stable conformations are required, separated by energy barriers. The height of these barriers should be several times larger than $k_\text{B}T$, to prevent thermal fluctuations from setting the molecule into random motion. To perform work, the transitions between the states should be unidirectional, which has so far been achieved in a few systems \cite{Koumura1999,Kelly1999,Leigh2003,Fletcher2005}. Most experiments to date have been performed on large assemblies of molecules \cite{Browne2006,Kay2007,Koumura1999,Kelly1999,Leigh2003,Fletcher2005}, although some examples have been studied where scanning tunneling microscopy was used to either manipulate or detect the motion on the single-molecule scale \cite{Zheng2004,Baber2008,Gao2008,Ye2010}.

\begin{figure}
    \begin{center}
        \includegraphics{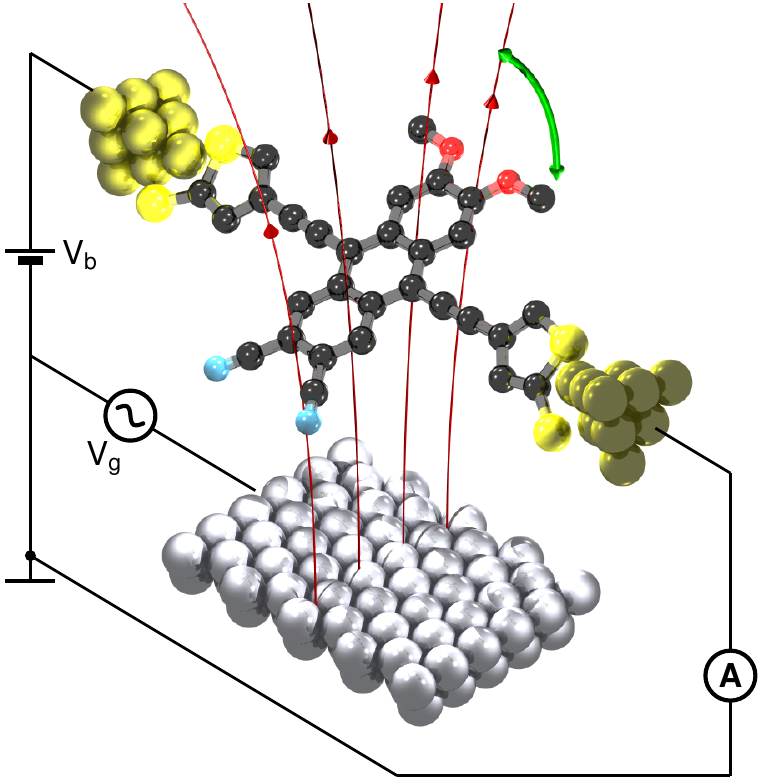}
    \end{center}
    \caption{\label{fig:design} Design of a molecular motor with a permanent electric dipole moment. The motor consists of anchoring groups connecting the conjugated backbone to the leads, allowing the measurement of the low-bias conductance, and a dipole rotor which can be driven by the oscillating gate field underneath. See also the animation in the Supporting Information.}
\end{figure}

In this paper, we propose a conceptually new design for a molecular motor, which enables the simultaneous driving and detection of the motion of a single-molecule motor at the nanoscale (see \ref{fig:design}). The rotating moiety is equipped with a permanent electric dipole moment and is part of a conjugated molecule, which is suspended between two metallic contacts above a gate electrode. By modulating the electric field generated by the gate, the dipole rotor can be driven to rotate with a speed controlled by the frequency of the gate field \cite{Horinek2005}. As it rotates, the rotor repeatedly switches between two stable states, each corresponding to a planar conformation where the molecule is fully conjugated. The unique aspect of this design is that by applying a small bias between the metallic contacts and measuring the current through the molecule, we can determine the position of the rotor, since a lowering of the conjugation during rotation has a dramatic effect on the conductance (see Methods).

\begin{figure}
    \begin{center}
        \includegraphics{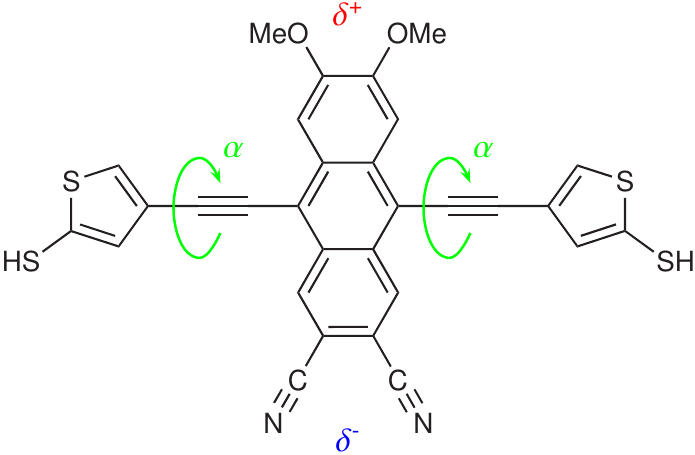}
    \end{center}
    \caption{\label{fig:molecule} Proposal for a molecular dipole motor: structure~\textbf{1} (9,10-bis((5-mercapto-3-thiophen-3-yl)ethynyl)-6,7-dimethoxyanthracene-2,3-dicarbonitrile). The blue $\delta^-$ and red $\delta^+$ symbols denote the partial charges inducing the dipole moment of the rotor, and the green arrow shows the axis facilitating a rotation of the rotor with an angle $\alpha$.}
\end{figure}

Structure~\textbf{1} (see \ref{fig:molecule}) is a simple example of a molecular dipole motor. It consists of three basic components: a dipole rotor, axles, and anchoring groups. The bidentate mercaptothiophene anchoring groups connect the molecule to the source and drain electrode, and are designed to provide good conductance and to limit conformational changes near the contact surface.\cite{Dulic2009} The ethynyl groups connected to the anchoring groups act as an axle about which the central anthracene moiety can rotate. The chemical design of these axles determines the height and shape of the rotational barrier potential. Finally, the anthracene rotor in between the axles interacts with an external electric field through the cyano ($\delta^-$) and methoxy ($\delta^+$) substituents, which induce a dipole moment ${\bm p}$. The dipole moment can be tuned by choosing different combinations of electron-withdrawing or electron-donating substituents and varying the distance between them.

We note that, in principle, it is also possible to use a magnetic field to drive a molecule with a magnetic dipole moment. However, the energy range accessible to a high-spin molecule ($S=5$) in a large magnetic field ($\left|{\bm B}\right|=10$~T) is limited to about $U=-{\bm\mu}\cdot{\bm B}=\pm 3$~meV, whereas the energy range of a molecule with a large \emph{electric} dipole moment ($\left|{\bm p}\right|=10$~D) in a large \emph{electric} field ($\left|{\bm E}\right|=1$~V~nm$^\text{-1}$) can be as large as $U=-{\bm p}\cdot{\bm E}=\pm 200$~meV. Moreover, electric fields can easily be applied locally in a molecular junction \emph{via} a gate electrode. Solving Poisson's equation for a typical geometry in an electromigrated break junction (EMBJ) with a gap separation of 2~nm yields an electric field of $\left|{\bm E}\right|\approx 1$~V~nm$^\text{-1}$ at the position of a molecular dipole due to the gate electrode for a gate range of $\pm5$~V. The calculated gate coupling at that position ($\beta=0.1$) corresponds well to typical values in measurements on single-molecule devices \cite{Osorio2010}.

\section*{Results and Discussion}

To analyze the behavior of the motor and to check whether its motion is detectable through current measurement, we have performed quantum chemistry calculations and classical Langevin simulations.

\begin{figure}
    \begin{center}
        \includegraphics{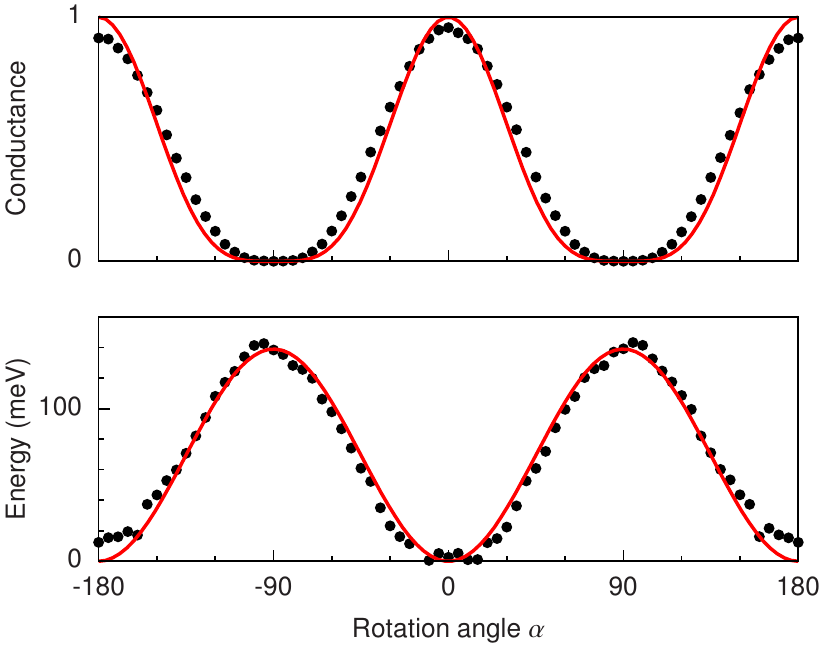}
    \end{center}
    \caption{\label{fig:potential} The rotational barrier potential (bottom) and normalized zero-bias off-resonance conductance (top) of structure~\textbf{1} obtained from DFT calculations as a function of the rotation angle $\alpha$ of the dipole rotor. The red lines illustrate the $\operatorname{sin}^2(\alpha)$ and $\operatorname{cos}^4(\alpha)$ behavior of the barrier potential and conductance, respectively.}
\end{figure}

\paragraph{Barrier Potential and Conductance.} Using density functional theory (DFT) and the Green's function method (see Methods), we calculate the rotational barrier potential and the zero-bias off-resonance conductance as a function of the rotation angle of the rotor ($\alpha$) and the anchoring groups ($\theta$), where $\alpha$ and $\theta$ are defined with respect to the direction of the electric field (vertical in the figures in this paper). For $\theta=0^\circ$, the results in \ref{fig:potential} show that for structure~\textbf{1} the barrier potential $U^r(\alpha)$ has a $\operatorname{sin}^2(\alpha-\theta)$ dependence, while the conductance is proportional to $\operatorname{cos}^4(\alpha-\theta)$. This strong dependence of the conductance on the rotation angle allows it to be used as a measure of the rotation. Note that the minima of the conductance correspond to the maxima of the barrier potential.

\paragraph{Driving and Dynamics.} For an electric field to rotate a dipole, the torque exerted by the field,
\begin{equation*}
\tau^p(\alpha)\equiv-\frac{\text{d}U^p(\alpha)}{\text{d}\alpha}=\left|{\bm p}\right|\left|{\bm E}(t)\right|\operatorname{sin}(\alpha),
\end{equation*}
should exceed the restoring torque in the molecule,\bibnote{Since the proposed design calls for a dipole close to metallic surfaces, one could in principle expect image charges to have an important effect on the potential landscape. However, as long as the distance between the gate and the center of the dipole rotor is larger than the diameter of the dipole, this has a negligible effect on the barrier potential (see Supporting Information).}
\begin{equation*}
\tau^r(\alpha)\equiv-\frac{\text{d}U^r(\alpha)}{\text{d}\alpha}=-U^r_0\operatorname{sin}(2(\alpha-\theta)),
\end{equation*}
for all values of $\alpha$ before the rotor crosses the maximum of the rotational barrier potential:
\begin{equation}
\left|{\bm p}\right|\left|{\bm E}\right|\operatorname{sin}(\alpha)\geq U^r_0\operatorname{sin}(2(\alpha-\theta)).
\end{equation}
In the case where $\theta=45^\circ$ the critical field is lowest ($\left|{\bm E}_c\right|=\frac{U^r_0}{\left|{\bm p}\right|}$), while for $\theta=0^\circ$ or $\theta=90^\circ$ it is twice as large.

Since thermal fluctuations are important, and can even be dominant for nanoscale devices, the dynamics of a molecular motor at temperatures above the quantum level splitting are most appropriately described by the Langevin equation \cite{Kubo1966,Gunsteren1982,Paterlini1998}:
\begin{equation}\label{eq:langevin}
I\frac{\text{d}^2\alpha(t)}{\text{d}t^2}=\tau^r(\alpha)+\tau^p(\alpha,t)-\gamma\frac{\text{d}\alpha(t)}{\text{d}t}+R(t),
\end{equation}
where $\alpha$ is the rotation angle of the dipole rotor, $I$ is the moment of inertia, $\gamma$ is a friction coefficient due to the coupling of the molecular motion to the phonon bath of the electrodes and the interaction between the dipole and the metallic surfaces \cite{Tomassone1997}, and $R(t)$ is a Gaussian distribution describing the thermal fluctuations, with a width of $2k_\text{B}T\gamma$ \cite{Gunsteren1982}.

In eq (\ref{eq:langevin}), the moment of inertia, height of the rotational energy barrier, and the dipole moment of structure~\textbf{1} are calculated using DFT, leaving only the friction coefficient $\gamma$ as a free parameter. In the limit of small oscillations around the potential minima, $\gamma$ enters into the solution of eq (\ref{eq:langevin}) as an exponential decay factor of $e^{-\frac{\gamma}{2I}t}$. This suggests defining $k_\nu=\frac{\gamma}{2I}$, where $k_\nu$ is the vibrational relaxation rate. For structure~\textbf{1}, we find $I=2.66\cdot10^7$~a.u. ($6.77\cdot10^\text{-44}$~kg~m$^\text{2}$), $U_0^r=139$~meV (or 3.21~kcal~mol$^\text{-1}$, corresponding to a torque of at most 2.23~$\times$~10$^\text{-20}$~Nm), and $\left|{\bm p}\right|=12.6$~D. The vibrational relaxation rate $k_\nu$ is chosen to be 10$^\text{9}$~Hz. Although it is possible to estimate certain contributions to the relaxation rate \cite{Tomassone1997}, this rather slow rate has been chosen because it represents a worst-case scenario from a device standpoint (slow relaxation amplifies thermal fluctuations). The results in this paper show, however, that even in this case the molecular motor can be driven and its motion can be measured.

\begin{figure*}
    \begin{center}
        \includegraphics{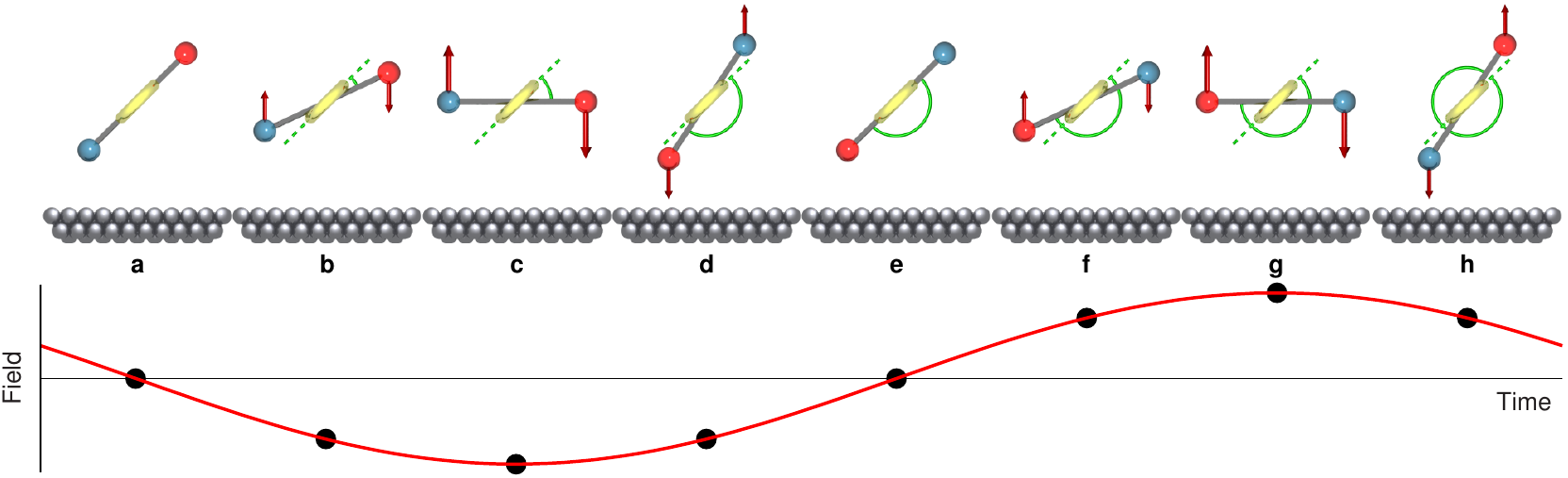}
    \end{center}
    \caption{\label{fig:dynamics} Dynamics of a dipole rotor in an electric field. (a)--(h) The position $\alpha$ of the dipole rotor as a function of time (top) due to an oscillating gate field (bottom). The rotor has been depicted schematically as seen along the rotation axis (see \ref{fig:design} for comparison), where the anchoring groups are shown in yellow, and the positive and negative partial charges of the dipole are shown in red and blue, respectively. When the field is zero, the rotor is in the equilibrium position (a) ($\alpha=45^\circ$). As soon as the field is turned on, the rotor feels a torque and starts to rotate (b). For this particular configuration, the torque becomes maximal at $\alpha=90^\circ$, precisely when the restoring torque in the molecule is also maximal (c). The rotor now accelerates and flips (d), ending up in the next equilibrium position when the field has vanished (e). The gate field then changes sign, and the process repeats (f)--(h).}
\end{figure*}

The position of a molecular dipole rotor as a function of an oscillating electric field is shown in \ref{fig:dynamics}. The anchoring groups are taken to bind under an angle of $\theta=45^\circ$ with respect to the direction of the field, as this results in the smallest gate voltage required to drive the motor (see Methods). It is clear from this figure that the motor is unidirectional, and can therefore perform work. This holds even when $\theta\neq 45^\circ$, although the critical field may then be larger (by at most a factor of two). The only two cases where the motor would not be unidirectional are $\theta=0^\circ$ and $\theta=90^\circ$ (within about $5^\circ$ at 15~K due to thermal fluctuations), as the rotor is then in the top (or bottom) dead center corresponding to the minimum or maximum of the barrier potential, where the direction of motion is random. Though unidirectional, the direction of motion will of course depend on the orientation of the molecule in the junction. For example, when $\theta=-45^\circ$ the direction of motion will be counterclockwise.

\begin{figure}
    \begin{center}
        \includegraphics{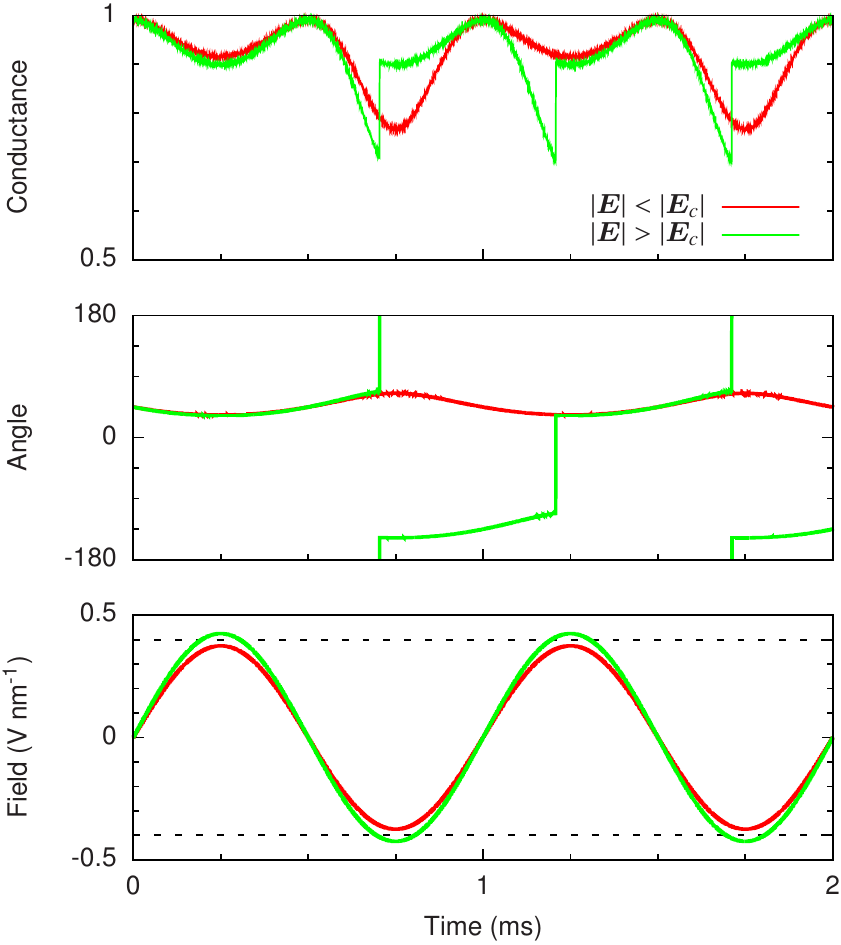}
    \end{center}
    \caption{\label{fig:conductance} The gate field, rotation angle, and normalized conductance as a function of time for structure~\textbf{1} in the configuration of \ref{fig:dynamics}. The red line corresponds to the situation where the field is too weak to drive the motor into rotation, and the green line to when it is just strong enough. For the first 0.5~ms the field is in the ``wrong'' direction, pushing the rotor counterclockwise instead of clockwise, and the conductance traces overlap since no rotation takes place. After that, the field is in the ``right'' direction, the motor starts to rotate, and the conductance traces start to deviate from each other. The dashed line indicates the amplitude of the critical field ($\sim$0.4~V~nm$^\text{-1}$). In the case of a rotating motor, the conductance has a period that is half that of the driving field and is no longer symmetric under time reversal (see \ref{fig:hysteresis}).}
\end{figure}

\begin{figure}
    \begin{center}
        \includegraphics{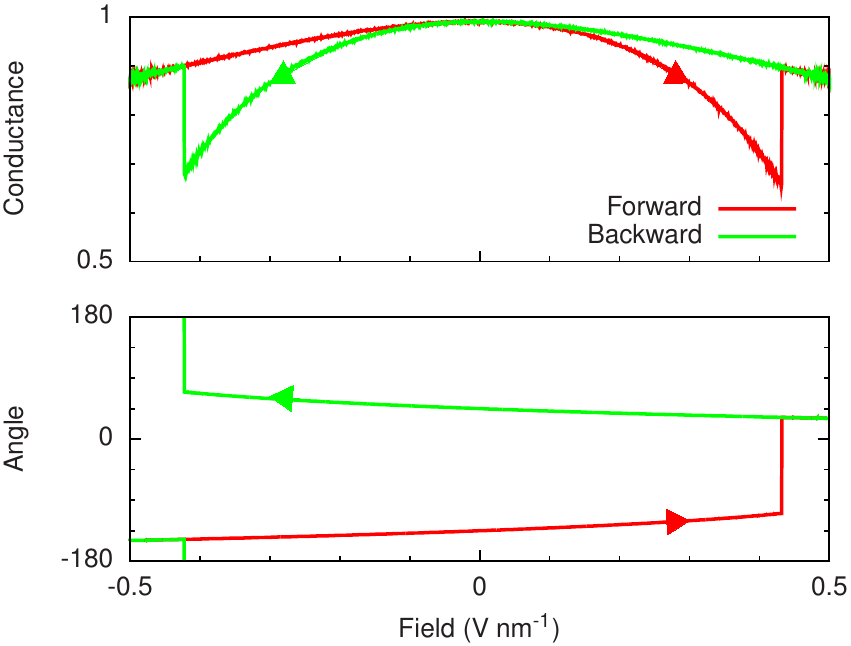}
    \end{center}
    \caption{\label{fig:hysteresis} The rotation angle and normalized conductance as a function of the applied gate field. The red line corresponds to the forward sweep of the field (negative to positive), and the blue line to the backward sweep. Both the rotation angle and the conductance clearly show hysteresis.}
\end{figure}

\paragraph{Detection of the Rotation.} To show how the rotational motion can be unambiguously detected, we have calculated the rotation angle $\alpha$ and conductance due to an oscillating gate field as a function of time. For $T=15$~K, the rotation angle and the conductance for structure~\textbf{1} are shown in \ref{fig:conductance} for a gate field with an amplitude just below (red line) and just above (green line) the critical value ($\sim$0.4~V~nm$^\text{-1}$ for $\theta=45^\circ$), which is well within the range accessible in current three-terminal devices. The rapid flip of the rotor (\ref{fig:dynamics}c,d) is visible as a vertical line in the rotation angle and as a switch in the conductance. While the potential changes with time, at any instance the rotor executes a Brownian motion around the equilibrium position, with the exception of the (nearly instantaneous) switch. The characteristics of \ref{fig:conductance} therefore do not change if the driving frequency is different, as long as it is lower than the relaxation rate ($>10^9$~Hz). This means that we have full control over the speed of motion, and therefore over the power output of the molecular motor, up to the GHz regime. At very low frequencies, the motor can even be driven statically as a switch.

Two features in the conductance plots in \ref{fig:conductance} enable us to distinguish between a rotating and a nonrotating molecular motor: the period and the time-reversal symmetry. While the period of the conductance is equal to that of the driving field for a merely oscillating motor, for a rotating motor it is half as long. More importantly, the conductance for a rotating motor is not symmetric under time-reversal. The reason for this can be seen in \ref{fig:hysteresis}, which shows hysteresis in the conductance as a function of the applied field: after the rotor flips, the field has to change sign for the rotor to return to its original position. Although the gate field may have an influence on the conductance besides the rotation of the dipole rotor, the time-reversal asymmetry in the conductance will still be present, and is therefore the hallmark of a molecular dipole motor.

\paragraph{Temperature Dependence.} The temperature of 15~K used in the calculation of \ref{fig:conductance} and \ref{fig:hysteresis} should be compared to the quantum-mechanical level splitting of the rotor vibrations ($\Delta E=\hbar\sqrt{2U^r_0/I}$, where $I$ is the moment of inertia). For the structure~\textbf{1} the level-splitting is 0.533~meV (6.19~K). As long as $k_\text{B}T$ is higher than the level splitting, the motor behaves classically, but at lower temperatures a quantum mechanical description becomes necessary. For example, at liquid helium temperatures, the motor will exhibit temperature-independent zero-point oscillations with an estimated amplitude of $2.5^\circ$. This leads to variations in the conductance which may be observable in the current noise.

\begin{figure}
    \begin{center}
        \includegraphics{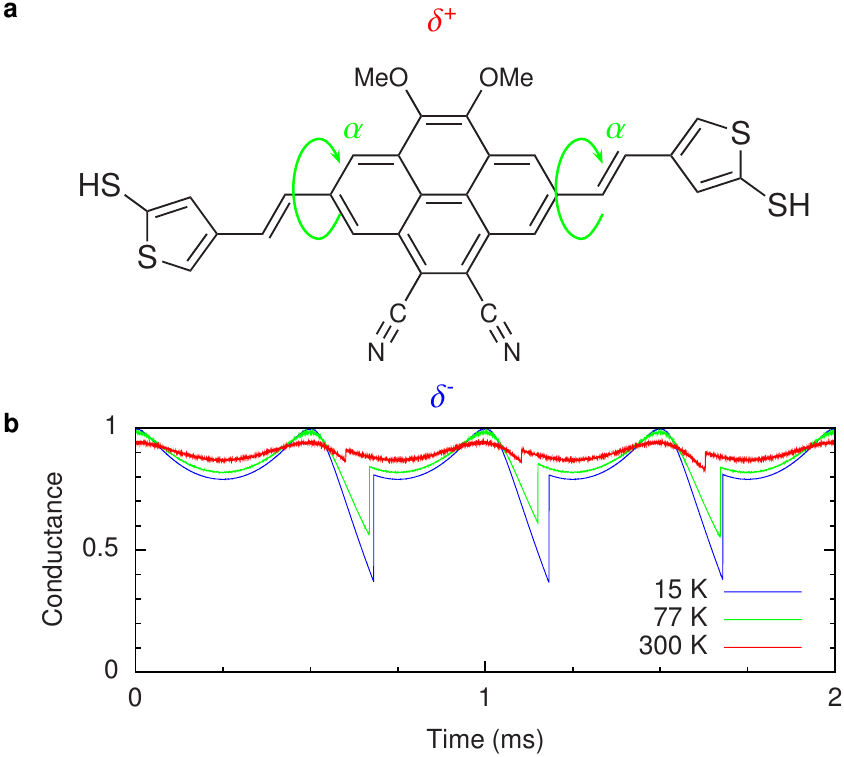}
    \end{center}
    \caption{\label{fig:temp} Design of a molecular dipole motor capable of room temperature operation. (a) Structure~\textbf{2} (2,7-bis((E)-2-(5-mercaptothiophen-3-yl)vinyl)-9,10-dimethoxypyrene-4,5-dicarbonitrile). The pyrene rotor has a dipole moment (${\bm p}$) of 9.8~D (b) Normalized conductance as a function of time at temperatures of 15, 77, and 300~K. The driving frequency is the same as in \ref{fig:conductance}, but the amplitude of the field is 1.75~V~nm$^\text{-1}$ at 300~K, 2.75~V~nm$^\text{-1}$ at 77~K, and 3.25~V~nm$^\text{-1}$ at 15~K.}
\end{figure}

At higher temperatures, thermal fluctuations cause the rotor to flip at a lower value of the gate field, making the difference between the conductance of the rotating and nonrotating motor less pronounced. Our calculations show that structure~\textbf{1} should still be operational at 77~K, but at room temperature it will rotate freely, even in the absence of a driving force. To allow operation at room temperature, a molecule with a higher rotational barrier potential is required. An example of such a design is shown in \ref{fig:temp}a (structure~\textbf{2}). Compared to structure~\textbf{1}, we have substituted the ethynyl groups in the axles by ethenyl and the anthracene rotor by a pyrene analogue. The axis of rotation is now located on the single bond between the ethenyl groups and the pyrene rotor. We have chosen pyrene instead of anthracene to minimize the steric hindrance between the rotor and the axles. These changes result in a 5-fold increase of the rotational barrier potential ($U_0^r=691$~meV or 15.9~kcal~mol$^\text{-1}$), preventing the motor from rotating freely at room temperature. The normalized conductance as a function of time for this molecule is shown in \ref{fig:temp}b for different temperatures. It is clear from this figure that the motor can still be driven and measured at temperatures up to 300~K, although the critical field at this temperature (1.75~V~nm$^\text{-1}$) may be challenging for current three-terminal devices.

\section*{Conclusions}

We have shown that the use of an electric field to drive a molecular dipole motor provides unidirectionality and complete control over the speed of rotation, while the conductance provides a real-time measure of the motion. The proposed molecule is easily synthesized and the parameters are such that it should be measurable in current electromigrated break junction setups. An important aspect of our design is the versatility offered by chemical synthesis. In particular, the barrier height, the dipole moment, and the moment of inertia of the rotor can all easily be changed. Our motor therefore constitutes a well-defined nanoelectromechanical system suitable for studying molecular motion over a wide range of temperatures, encompassing both the classical and quantum regime.

\section*{Methods}

\paragraph{Conductance as a Function of Angle.} As shown in \ref{fig:potential}, the rotation of the dipole rotor can be determined by measuring the low-bias off-resonance conductance $G(\alpha)$ of the motor. In the coherent transport regime, this conductance is given by \cite{Datta1995}
\begin{equation}
G=g_0\operatorname{Tr}\left\{\hat{\Gamma}_\text{L}(E_\text{F})\hat{G}^-(E_\text{F})\hat{\Gamma}_\text{R}(E_\text{F})\hat{G}^+(E_\text{F})\right\},
\end{equation}
where $g_0$ is the conductance quantum, $\hat{\Gamma}_\text{L}$ and $\hat{\Gamma}_\text{R}$ are the imaginary parts of the self-energy matrices associated with the left and right electrode, and $\hat{G}^-(E)=\left(E-H-\text{i}\epsilon^+\right)^{-1}$ is the retarded Green's function. For simplicity, we have not explicitly included the contacts in the calculation of the Green's function. $\hat{\Gamma}_\text{L}$ and $\hat{\Gamma}_\text{R}$ are assumed to couple only to the sulfur atoms in the anchoring groups, and the Fermi energy $E_\text{F}$ is chosen to be 0.5~eV from the highest occupied molecular orbital (HOMO). Choosing a different value does not significantly alter the results. It is known that a gate field can influence the conductance by shifting the HOMO and lowest unoccupied molecular orbital (LUMO) with respect to the Fermi energy, but since the molecule is strongly coupled to the leads, this effect is expected to be much smaller than the change in conductance due to the lowering of the conjugation. Although a full NEGF calculation can certainly improve on these results, this approach semiquantitatively captures the behavior of the conductance as a function of the rotation angle.

For a conjugated molecule, conductance takes place primarily \emph{via} the HOMO or LUMO, which consist of a hybridization of the p$_\text{z}$-orbitals on the carbon and sulfur atoms of the backbone. Since the dipole rotates around the ethynyl ``axles'', we expect the conductance to be proportional to the overlap between the p$_\text{z}$-orbitals on the backbone. To first order, this overlap is proportional to $\operatorname{cos}^2(\alpha-\theta)$, which has also been observed in measurements \cite{Venkataraman2006,Vonlanthen2009,Mishchenko2010}. Therefore, since a dipole motor has two axles, the conductance is expected to be proportional to $\operatorname{cos}^4(\alpha-\theta)$, or, when $\theta_l\neq\theta_r$:
\begin{equation}\label{eq:conductance}
G(\alpha)\sim\operatorname{cos}^2(\alpha-\theta_l)\cdot\operatorname{cos}^2(\alpha-\theta_r),
\end{equation}
where each axle contributes a factor of $\operatorname{cos}^2(\alpha)$ to the conductance. Since the peaks in $G(\alpha)$ become narrower when $\theta_l\neq\theta_r$, the conductance is slightly more sensitive to the rotation angle. However, the overall behavior of the potential and the conductance do not change when $\theta_l\neq\theta_r$ and the proposed actuation and detection principle should work for a large variety of experimental configurations. Plots of the barrier potential and the conductance of structure~\textbf{1} for $\theta_l-\theta_r=45^\circ$ are given in the Supporting Information.

All quantum chemistry calculations have been performed with the Amsterdam Density Functional package \cite{Guerra1998,Velde2001}, using the PW91 exchange-correlation potential and a triple-$\zeta$ doubly polarized basis set.

\paragraph*{Acknowledgment.} We thank M.~Poot, J.~van~Esch, H.~Valkenier, and Y.~Blanter for discussions. Financial support was obtained from Stichting FOM (project 86), and from the EU FP7 program under the grant agreement ``SINGLE'.'

\paragraph*{Supporting Information Available:} Analysis of image charge effects, the angle dependence of the barrier potential and the conductance, and an alternative design for a molecular dipole motor, as well as animation of the rotation of a dipole rotor in an oscillating electric field. This material is available free of charge \emph{via} the Internet at http://pubs.acs.org.

\providecommand*{\mcitethebibliography}{\thebibliography}
\csname @ifundefined\endcsname{endmcitethebibliography}
{\let\endmcitethebibliography\endthebibliography}{}


\begin{mcitethebibliography}{27}
\bibitem[Not()]{Note-1}
 Since the proposed design calls for a dipole close to metallic surfaces, one
  could in principle expect image charges to have an important effect on the
  potential landscape. However, as long as the distance between the gate and
  the center of the dipole rotor is larger than the diameter of the dipole,
  this has a negligible effect on the barrier potential (see Supporting
  Information).\relax
\mciteBstWouldAddEndPunctfalse
\mciteSetBstMidEndSepPunct{\mcitedefaultmidpunct}
{}{\mcitedefaultseppunct}\relax
\EndOfBibitem
\end{mcitethebibliography}
\end{document}

\section*{Image charge effects}

\begin{figure}
    \begin{center}
        \includegraphics[scale=2]{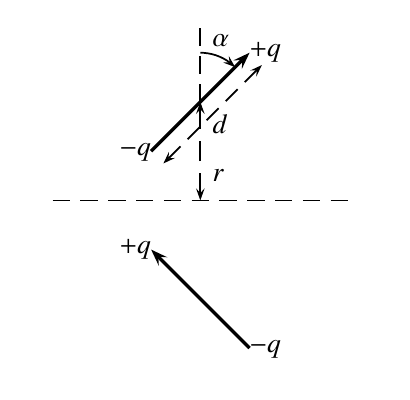}
    \end{center}
    \caption{\label{fig:image} An electric dipole near a dielectric surface (dashed line), with the corresponding image dipole. $r$ is the distance between the center of the dipole and the surface, while $d$ is the diameter of the dipole itself.}
\end{figure}

A schematic picture of an electric dipole near a dielectric surface, with the corresponding image dipole, is shown in Fig.~\ref{fig:image}. Defining $\epsilon=\frac{d}{2r}$ and $\theta=\operatorname{arctan}(\epsilon\operatorname{sin}(\alpha))$, the torque on the dipole due to static interactions with the image dipole is given by
\begin{equation}
\tau=-\frac{p^2}{4\pi\epsilon_0}\frac{1}{8r^3}\left(\frac{\operatorname{sin}(2\alpha)}{\left(1-\epsilon^2\operatorname{cos}^2(\alpha)\right)^2}-\frac{\operatorname{cos}(\alpha)\operatorname{sin}(\theta)}{\epsilon\left(1+\epsilon^2\operatorname{sin}^2(\alpha)\right)}\right).
\end{equation}
In the limit of $\epsilon\rightarrow 0$, this reduces to
\begin{equation}
\tau=-\frac{p^2}{4\pi\epsilon_0}\frac{\operatorname{sin}(2\alpha)}{16r^3}.
\end{equation}
Since $\tau\equiv-\frac{\text{d}U}{\text{d}\alpha}$, the angular dipole image potential is
\begin{equation}
U=U_0\operatorname{sin}^2(\alpha),
\end{equation}
where
\begin{equation}
U_0=\frac{p^2}{4\pi\epsilon_0}\frac{1}{16r^3}.
\end{equation}
In this limit, for a system with $p=10$~D and $r=1$~nm, $U_0\approx 3.90$~meV. For most molecular motors this effect will be negligible, since the rotational barrier potential typically exceeds 100~meV.

\section*{Angle dependence of $U^r$ and $G$}

\begin{figure}
    \begin{center}
        \includegraphics{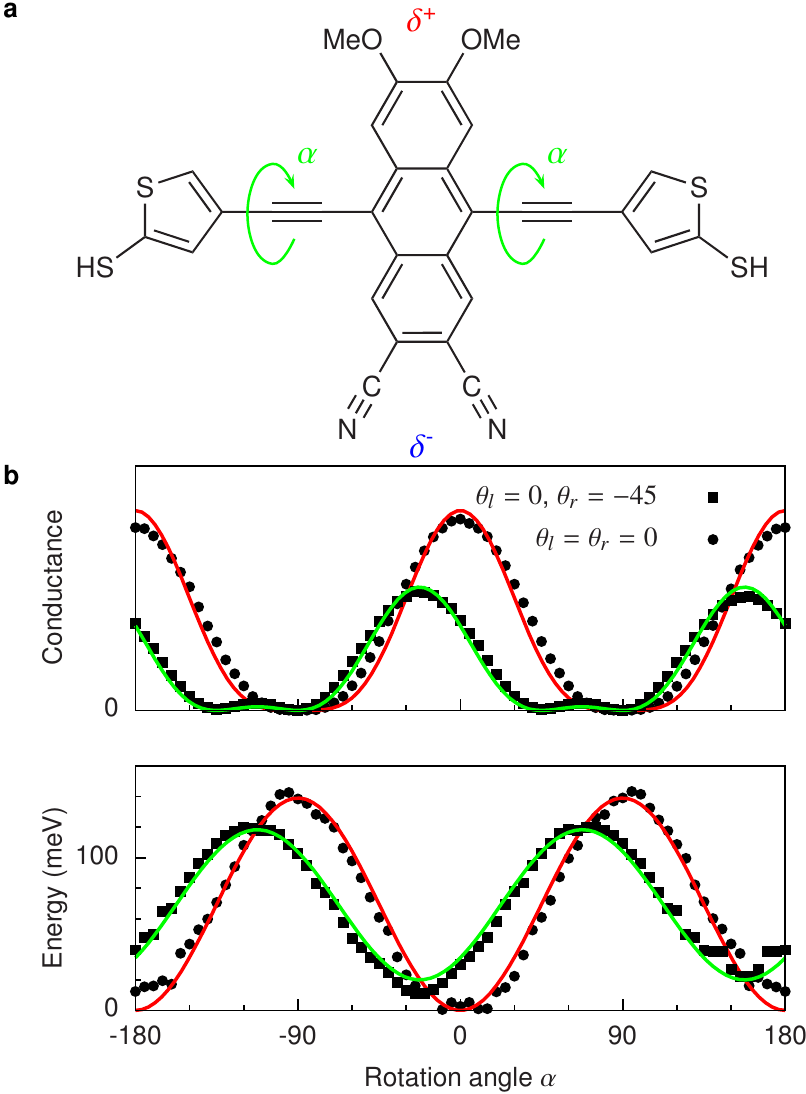}
    \end{center}
    \caption{\label{fig:fit} Potential and conductance as a function of the rotation angle. (a) Structure~\textbf{1}. (b) The rotational barrier potential (bottom) and normalized zero-bias off-resonance conductance (top) obtained from DFT as a function of the rotation angle $\alpha$ of the dipole rotor, in the case where $\theta_l=\theta_r=0$ and $\theta_l=0$, $\theta_r=-45$. The red and green lines show the fits to eqs (\ref{eq:potential}) and (\ref{eq:conductance}) of the barrier potential and conductance, respectively.}
\end{figure}

As shown in Figure~3 in the paper, the rotational barrier potential $U^r(\alpha)$ of a molecular dipole motor has a $\operatorname{sin}^2(\alpha-\theta)$ dependence on the rotation angle $\alpha$ of the rotor, where $\theta$ is the equilibrium angle. Since a conjugated molecule prefers to be planar, $\theta$ is the angle under which the anchoring groups bind to the surface. However, in a junction the binding angles of the left and right anchoring groups might be different, \emph{i.e.}, $\theta_l\neq\theta_r$. This shifts the minima of the potential and lowers the barriers between them, but it does not change the sinusoidal behavior of the potential:
\begin{equation}\label{eq:potential}
U^r=U^r_0\operatorname{cos}(\theta_l-\theta_r)\operatorname{sin}^2\left(\alpha-\frac{\theta_l+\theta_r}{2}\right).
\end{equation}
As shown in the methods section in the paper, the conductance changes according to
\begin{equation}\label{eq:conductance}
G(\alpha)\sim\operatorname{cos}^2(\alpha-\theta_l)\cdot\operatorname{cos}^2(\alpha-\theta_r).
\end{equation}
This can be clearly seen in \ref{fig:fit}, where the barrier potential and the conductance obtained from DFT have been plotted in the case where $\theta_l=\theta_r=0$ and in the case where $\theta_l=0^\circ$, $\theta_r=-45^\circ$. Both plots of the potential have been fitted with eq (\ref{eq:potential}) using the same value for $U^r_0$ (139~meV).

\section*{Alternative design}

\begin{figure}
    \begin{center}
        \includegraphics{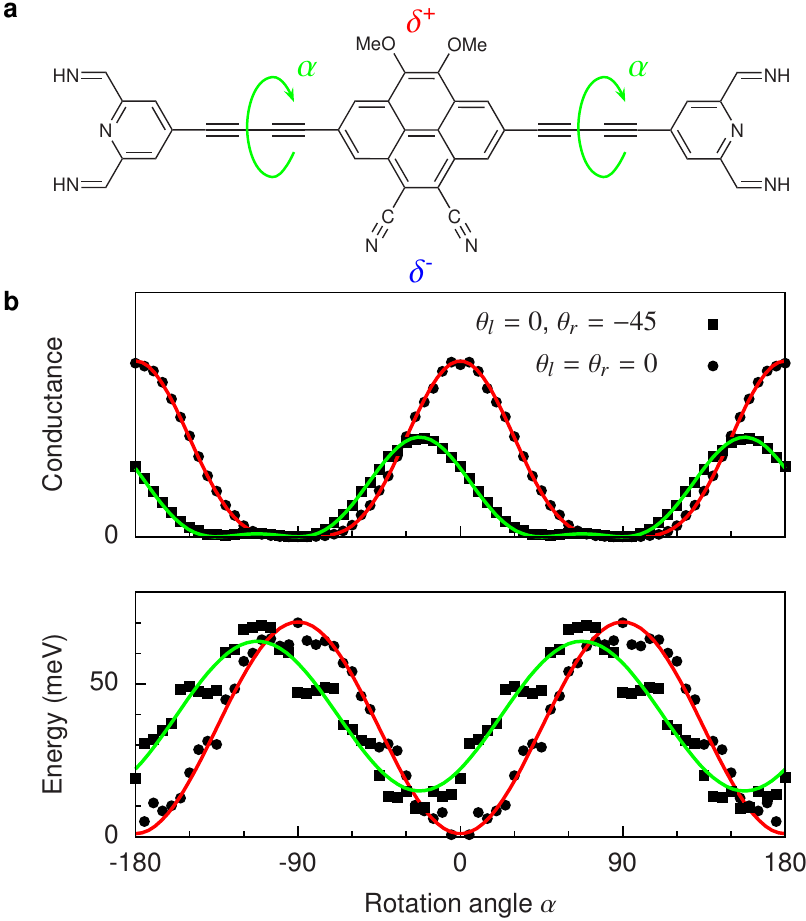}
    \end{center}
    \caption{\label{fig:molecule} Alternative design for a molecular dipole motor. (a) Structure~\textbf{3} (2,7-bis((2,6-bis(iminomethyl)pyridin-4-yl)buta-1,3-diynyl)-9,10-dimethoxypyrene-4,5-dicarbonitrile). Instead of ethynyl, this design uses diynyl groups as axles, lowering the rotational barrier to 69~meV (1.59~kcal~mol$^\text{-1}$). The smaller distance between the cyano and methoxy anchoring groups on the pyrene rotor compared to athracene results in a dipole moment of 9.8~D. (b) The rotational barrier potential (bottom) and normalized zero-bias off-resonance conductance (top) obtained from DFT as a function of the rotation angle $\alpha$ of the dipole rotor. The red and green lines show the fits to eqs (\ref{eq:potential}) and (\ref{eq:conductance}) of the barrier potential and conductance, respectively.}
\end{figure}

A possible alternative design for a molecular dipole motor is shown in ~\ref{fig:molecule}a (structure~\textbf{3}). The anchoring groups and axles are different compared to structures \textbf{1} and \textbf{2} in the paper, but the qualitative behavior remains the same. As can be seen in ~\ref{fig:molecule}b, the potential and conductance again fit quite well to eq (\ref{eq:potential}) and (\ref{eq:conductance}).

The increased length of the molecule (2.9~nm instead of 1.8~nm) will make it easier to trap the molecule in a breakjunction, while the fact that the pyrene rotor (the same as in structure~\textbf{2}) does not extend as far from the axis of rotation as anthracene will make it less likely for the rotation to be obstructed by the environment. Despite the lower dipole moment of the pyerene rotor of 9.8~D (compared to 12.6~D for structure~\textbf{1}), the reduction in the barrier height to 69~meV (1.59~kcal~mol$^\text{-1}$) due to the diynyl axles results in an overall reduction of the critical field by almost 40\%. The lower barrier potential may facilitate experiments designed to study molecular motion in the quantum regime. On the other hand, this design may put higher demands on the synthesis.